\documentstyle[12pt,epsf,graphicx]{article}

\begin{document}

\title{A Quark-Matter Dominated Universe}
\author{D.~Enstr\"{o}m, S.~Fredriksson, J.~Hansson
\\ Department of Physics \\
Lule\aa \ University of Technology\\
SE-971 87 Lule\aa , Sweden
\vspace{3mm}
\and A.~Nicolaidis
\\ Department of Theoretical Physics \\
Aristotle University of Thessaloniki \\
GR-540 06 Thessaloniki, Greece
\vspace{3mm}
\and S.~Ekelin
\\ Department of Mathematics \\
Royal Institute of Technology \\
SE-100 44 Stockholm, Sweden}

\date{}

\maketitle

\begin{abstract}
We present a new scenario for the development
of the Universe after the Big Bang, built on the
conjecture that a vast majority of the primordial
quark matter did not undergo a phase transition
to normal nuclear matter, but rather split up
into massive quark objects that remained stable.
Hence, such primordial quark matter would make up
the so-called dark matter. We discuss, mostly in
qualitative terms, the consequences for galaxy
formation, the origin of normal matter, the occurrence
of massive black-holes in galactic centres and the
cosmic gamma-ray bursts.
\end{abstract}

\section{Introduction}

One of the most fascinating mysteries in modern
astrophysics and cosmology is the nature and origin
of the so-called dark matter in the cosmos. It is
(by definition) non-luminous, and reveals
itself only through its gravitational interaction
with the luminous galactic matter or with
light. Its main signature is that most studied
galaxies rotate in a ``non-Keplerian'' way,
as estimated from their luminous matter
\cite{Raffelt97}. It appears as if the galaxies
contain some extra, non-luminous matter,
with a total mass believed to be
about an order of magnitude higher than
that of the luminous matter. There are weaker
indications that also galaxy clusters
behave strangely, and therefore would
contain some extra matter inbetween the
galaxies. There are also theoretical,
cosmological arguments favouring an
overall densitity much higher than the
one estimated from direct observations
of luminous matter.

Detailed studies of galaxy rotations seem to indicate
that dark matter has a more extended density
profile than the luminous matter, stretching out to
several times the conventional galactic radii.
However, there is no general agreement as to the
geometrical shape of the dark-matter halo
of a typical spiral galaxy, and suggestions of
spherical as well as slightly flattened mass
distributions can be found in the literature.
Neither is there a generally accepted functional
dependence of the density $\rho (r)$ as a function
of the distance $r$ to the galactic centre, even
for fits to spherical distributions. One often assumes
a form that at least asymptotically falls off
as $r^{-\alpha}$. In most analyses an
$\alpha < 2$ is used, so that one has to introduce
a cut-off in $r$ in order to avoid an
infinite galactic mass; {\it i.e.},
an {\it ad hoc} galactic radius.
Normally, one assumes that the total
mass of a galaxy is around ten times
that of the luminous matter, while the
radius can vary widely, depending on the
exact choice of density distribution.
Some recent discussions of best-fit analytical
forms can be found in \cite{Moore97}.
The widely varying forms originate from different
analytic methods (except for the trivial reason
that they are sometimes used for different galaxies).
Different groups aim at fitting different parts of
the dark-matter distribution, which sometimes refers to
all galactic dark matter, and sometimes only to the
part being more peripheral than our solar system.
There is also a distinction between analyses
that are founded on model-dependent
simulations of galaxy formation, including
conjectured values of cosmological parameters,
and those built on observations of the Milky Way
and other galaxies.

During the last few years an extensive study of stars
in the Large Magellanic Cloud (LMC) has also found a few
cases of gravitational lenses, so-called machos
(Massive Compact Halo Objects), which magnify the
light from those stars, and seem to move in the
outskirts of our galaxy. The masses of the discovered
objects lie below the solar mass ($M_{\odot}$),
maybe in the range $(0.01-0.8) M_{\odot}$
\cite{Alcock96}. It is an open
question if the machos can make up for all
dark matter in the Milky Way.
The MACHO collaboration itself gives some
support to an average value for the macho mass
of $0.5M_{\odot}$, and a $50$ {\it per cent}
macho fraction in the Milky Way halo
(assuming a spherical shape).

There is no lack of imaginative models for the
dark matter, a majority of which rely on
ideas that have never been confirmed, or even
supported, by independent experiments or
measurements on earth or in space \cite{Raffelt97}.
The least speculative ones are those built on
astrophysical ideas about dark, compact objects
of normal matter, {\it e.g.}, bodies created
like normal stars, but with too small masses
to ignite fusion processes and become luminous.
So-called jupiters and brown dwarfs fall into
this category. One can also think of planets or
comets that have escaped solar systems in large
numbers.

Most explanations built on particle physics
are much more speculative. Sometimes they even rely
on new fundamental ideas that were invented
{\it just} to explain dark matter. Examples of
interesting, but entirely hypothetical, particles
assumed to contribute to the dark matter,
are ``neutralinos'' and ``axions''
\cite{Raffelt97}.

There are only two particle-physics
motivated dark-matter
models that are based on relatively well-known,
or theoretically well-studied phenomena.
The simplest one is the ``heavy-neutrino''
model, {\it i.e.}, the suggestion that at least one
neutrino species has a rest mass high enough to
make up for the galactic dark matter.
Such neutrinos cannot, of course, explain the discovered
gravitational lenses in the Milky Way, neither can they
easily be reconciled with the fact that dark matter
seems more peripheral than normal matter. There
are also more general problems with galaxy
formation, and it
is believed that heavy neutrinos
can account for only a small fraction of
the dark matter \cite{Sarkar97}.

The other model of this kind identifies dark matter
with objects consisting of a so-called {\it quark-gluon
plasma} (QGP), {\it i.e.}, a form of matter with only
quarks, and no structuring into protons and neutrons.
Such objects are expected from, or at least not
forbidden by, basic quark theory; quantum
chromodynamics (QCD). There is no reason to
believe that systems of just a few quarks
(two or three) would be the only ones of physical or
astrophysical interest. There is indeed an intense
current research about QGPs.
One example is the extensive experimental
programmes at several high-energy laboratories,
which aim at creating a QGP in heavy-ion
collisions.
The idea is to compress nuclear matter into such a
dense state that individual nucleons can no longer be
distinguished. Then the QGP might perhaps survive
long enough to send out some clear signals, before
converting (``hadronising'') into normal matter again.
A few hints of QGP creation have indeed been found
(``strangeness enhancement'' and ``$J/ \Psi$ suppression''),
above all at the CERN laboratory in Geneva,
but these have also
been disputed, and claimed to be consequences
of more conventional physical phenomena.

The extension of QGP ideas into astrophysics
is straightforward, since the bulk
of matter must have been in the form of
such a plasma just after the Big Bang, when
densities were still far above those inside atomic
nuclei, and of the order of those inside protons and
neutrons. There are also many current astrophysical
situations where one can think of extremely high
densities, such as inside a neutron star, or a
collapsing would-be supernova.

Normal hadronic matter is believed to have been created
spontaneously as soon as the Universe expanded into
an average density of the order of normal nuclear
densities, although there is no agreement as for the
details of this universal hadronisation. One can think
of an explosive and very fast phase transition, going
from the cooler periphery of the Universe and
inwards, or a slower growth of bubbles of normal
matter inside local density fluctuations, until finally
the QGP disappeared, as shrinking
bubbles inside the normal hadronic matter.

It seems a very natural line of thinking to speculate
that something went wrong within this scenario,
so that a vast majority of the primordial matter stayed
in the QGP phase, and now constitutes the dark
matter. From a minimalistic point of view it is more
natural to build on the fact that the Universe
has {\it already been} in a specific ``dark-matter'' state,
than to speculate about completely unknown forms
of matter.

However, this requires the QGP to be {\it the absolute
ground state of matter}, at least in some cosmically
interesting mass region. Ideas along such lines
began to flourish in the early 1970s, with a pioneering
work by Bodmer \cite{Bodmer71}, and subsequent
analyses in 1979 by Chin and Kerman \cite{Chin79},
and Bjorken and McLerran \cite{Bjorken79}.
In 1984 De R\'ujula and Glashow \cite{Rujula84},
Fahri and Jaffe \cite{Fahri84}, and
Witten \cite{Witten84} presented more refined
analyses, which strengthened the conclusion that
quark-matter ``nuggets'' or ```bags'' are indeed the
ground state of matter, and hence
contribute to dark matter.

This idea cannot be rigorously underbuilt by basic
principles, {\it i.e.}, from QCD.
Rather, the rule of the game is to rely on
``QCD-inspired'' phenomenological models, and
the one most frequently used for analysing existing
and hypothetical multi-quark objects is the so-called
MIT bag model \cite{Chodos74}. It is built
on the assumption that
quarks are confined to hadronic ``bags'' due to an
external ``vacuum'' pressure, quantified by the
so-called bag constant $B$, which takes a universal
value, normally given as $B^{1/4} \approx 150$
MeV. The $B$ value is fitted to known properties
of normal hadrons, in a variational procedure, where
the total energy (mass) of a hadron, primarily the
proton, is minimised with the help of the bag radius.
The known values of the proton mass and radius
are then used to fit the model parameters, $B$ being
the most crucial one. Many versions of the MIT bag
model have been developed, containing various
corrections, one of which is for interactions among
the quarks, which were assumed to be free inside the
bag in the original version.

When analysing bags with more than three quarks,
it appears as if already those with six quarks have
a chance of being stable against decays via strong
interactions. The case of such ``$H$ dibaryons''
\cite{Jaffe77} is still debated in the literature, but no
experimental evidence has so far been found.
There is a clear trend within the model
toward a higher stability for heavier objects with
even more quarks. Such a stability appears in the
analysis as a lower total energy per quark than
inside a single nucleon. However, this conclusion
requires that heavy quark objects, and already the
$H$ dibaryon, contain an equal (or almost equal)
number of the three lightest quarks, the $u$, the $d$
and the $s$ quarks. In this respect, the quark objects
predicted by the MIT bag model differ from
the structure of a hypothetical, strongly compressed
atomic nucleus. The reason that the $s$ quarks
are so crucial, in spite of their assumed higher mass,
is that the Pauli principle does not forbid them to
occupy the low-energy states of the $u$ and $d$
quarks. Once the $s$ quarks are captured in low-lying
energy levels they cannot decay through weak
interaction, much in the same way as neutrons in
atomic nuclei can be stable.

Still, some authors present results for a QGP of just
$u$ and $d$ quarks, but this is mainly motivated with
simplicity arguments in order to avoid computational
difficulties.

Hence, the macroscopic quark objects predicted to be
of astrophysical value are often referred to as
strangelets, strange stars, etc. There are several
obvious, but interesting, consequences of this peculiar
composition. One is that a quark object can be
electrically neutral, due to the charges
$(+2e/3,-e/3,-e/3)$ of the $(u,d,s)$ quarks. Hence
a ``strangelet'' qualifies as dark matter, since it cannot
attract an electron cloud, and therefore not send out
light through ``atomic'' processes. It can still emit
temperature radiation would it have a hot surface,
but if it represents the ground state of matter, it will
do so even at temperature $T = 0$.

Another observation is that a strangelet cannot easily
be produced by a contraction of atomic nuclei
in high-energy heavy-ion reactions at
accelerators. It would require weak-interaction
conversions to $s$ quarks of many of the original
$u$ and $d$ quarks in the very short time available
before the nuclear matter flies apart again. At the
best, one can hope for a ``baryon-free''
plasma, consisting of newly created quarks
and antiquarks. If so, one might detect an
enhanced production of
hadrons containing $s$ and $\bar{s}$ quarks,
as the result of a
limited QGP creation. The experimental evidence
for enhanced strangeness due to such processes is
under an intense study \cite{Greiner98}.

However, an almost exact balance between the three
quark species is expected inside the cosmological
plasma created after the Big Bang. In fact, ``any''
set of equally fundamental particles were
in balance as long as the
temperature (given in conventional mass units) was
still much higher than the particle masses. Therefore,
it suffices that the temperature obeyed
$T \leq 150$ MeV at
the time of the $s$ quarks being captured inside
their final quark objects. More massive quarks
($c$, $b$ and $t$) are not believed to exist in
significant numbers inside absolutely stable quark
matter, although they certainly contributed to the
processes in the very early Universe.

In the early work on such primordial quark nuggets
in the cosmos, the authors did not commit themselves to
a certain size (mass) of the objects, nor to an estimate
of their absolute importance as dark matter.
This probably has
to do with the fact that numerical results from the
MIT bag model, for objects containing more than a
handful of quarks, are unreliable. The original results
for protons and other low-mass hadrons were
achieved through exact solutions of the Dirac equation
inside a spherical bag. Various approximations must
be applied for systems with dozens of quarks, not to
mention the $10^{30}$ quarks, or more, inside
astrophysical objects. Many of these are
built on computational methods from nuclear physics,
or from relativistic statistical mechanics.

One typical MIT bag model analysis of this kind,
presented in \cite{Greiner98}, shows that a
QGP with less than about $20$ quarks has a
higher total energy than a corresponding nucleus,
or set of free nucleons. This result assures that light
nuclei do not decay to a QGP. However, heavier objects
have lower total energies per quark, and are
hence stable according to this model. This
has been analysed up to a few hundred
quarks, where the limit is set by practical
computer capacities. We will assume that these
MIT bag model results apply all the way up to
quite heavy objects. For extremely heavy
(astrophysical) objects also gravity is assumed
to play an important role, and the MIT bag
model must be complemented with general
relativity. We will discuss this problem later.

Strangely enough, this simple and logical dark-matter
model does not seem to have reached a popularity in
line with those built on much more exotic and
controversial ideas. The modern literature on a
quark-object dominated Universe is scarce, and, in
fact, non-existent (would it not be for a persistent
interest in its very early stage, and the
presumed phase-transition from quark-matter
to hadrons).

This neglect cannot have been caused by published
counter-arguments, because these are scarce
too. A much quoted one is an analysis by Madsen
\cite{Madsen88}. He studied the effect of quark nuggets
in space hitting a neutron star, and found that an
impact of a very small amount of strange QGP into
the dense interior of a normal neutron star would
catalyse a phase transition of its full core. Hence,
even a tiny lump of quark matter would turn a
neutron star into a ``quark star'' (or ``hybrid star''),
with a QGP-dominated interior. Such an extremely
compact core would prevent the neutron star from
experiencing a so-called pulsar glitch, {\it i.e.},
a sudden change of rotational frequency, which is
believed to be caused by a ``starquake'' coupled to
an immediate change of the moment of inertia around
the rotational axis. Since the probability of a pulsar
glitch occurring in a neutron star can be estimated
from observations, these arguments lead to an upper
limit for the density of quark-matter nuggets in space,
which, according to Madsen, excludes them
as the dominant dark matter.

However, Madsen's conclusion has been disputed, and
there are even claims that pulsar glitches {\it require}
the presence of a quark-matter core, or are caused by
the very phase transition when the core turns from
neutron to quark matter. Hence, there is by now a rich
literature built on the idea that some, or all, neutron
stars have quark-matter cores. In fact, the main
interest in cosmic quark matter in the modern
literature now focuses on such conventional compact
stars. Here, the quark matter is assumed to appear as
the result of a gravitational contraction of normal
matter, or of a rapid collapse after a supernova
explosion. A review of the current literature can be
found in Glendenning's book {\it Compact Stars}
\cite{Glendenning96}.

The true reason for quark matter to be out of fashion
as a dark-matter candidate therefore seems to be
sociological, namely that other ideas are more in line
with the current development in theoretical and
experimental high-energy physics. Here, the trends
are toward ``smaller length-scales'' and higher
mass-scales, parallel to the construction of new
accelerators at CERN and elsewhere. This has led to
a strongly increased interest in concepts like
supersymmetry, leptoquarks, massive neutrinos and
Higgs bosons, including a full spectrum of interesting
astrophysical and cosmological implications. An
important reason is probably also that some of these
ideas, {\it e.g.}, on massive neutrinos and on
supersymmetric partners of quarks and leptons, can
be tested with the impressive astrophysical Cherenkov
detectors now in use around the globe.
Dark-matter quark objects
would be more elusive in this sense, primarily
because they would be orders of magnitude less
frequent than the exotic single-particle candidates,
due to their much higher masses. Also, the atmosphere
(or the water, ice or rock surrounding the Cherenkov
detectors)
would presumably erase their traces. Such traces are
not even well-defined, and can therefore probably not
be discriminated from an impact of normal atomic
nuclei in, {\it e.g.}, a space-born detector.

Nevertheless, we argue that the original ideas
of quark objects as dark matter are worthy of
a revival, as they have not been convincingly
counter-proved, and since they are built on
a very simple principle, {\it i.e.}, the one about
the absolute stability of massive multi-quark
states. We will assume that practically all
dark matter is in this form, and try to
limit the model parameters with the help
of astrophysical data. For natural
reasons, the discussion will be mainly
qualitative since really conclusive data
are indeed scarce. Also, the particle-physics
foundations are not too well known, and
in particular not the proper way of
using the MIT bag model for very
massive objects.

\section{Quark objects in the early Universe}

\subsection{General considerations}

We take it for granted that a quark-gluon
plasma (even at $T = 0$) represents the absolute
ground state of matter, at least beyond some minimal
number of quarks, as indicated by the MIT bag model
results. This limit can be of the order of a few dozen
quarks, without being in conflict with the stability of
normal matter, since a spontaneous decay of a
heavy nucleus into a QGP would be exceedingly
improbable. The most natural way for
matter to stabilise after the most violent expansion
phase following the Big Bang would then
be to remain as massive quark objects.
Hence, the global plasma simply split up into
smaller objects in a more or less random way
due to the expansion after the quarks had been
created, and after the matter-antimatter
annihilation period had ended.

Such a split-up is probably impossible to analyse
in quantitative terms, regarding all uncertainties in
the standard Big Bang scenario. The consequence
crucial for the further development is naturally
the distribution of quark-object masses and radii after the
split-up phase. A separation of two regions of
quark matter could be a consequence
of the strong quark-confining forces and to some
extent gravitation, both counteracting
the general expansion. One can also think
of an earlier cause of this phenomenon,
{\it e.g.}, through very early quantum-mechanical
fluctuations or a split-up before (hypothetical)
subquarks joined to form the quarks. The split-up would
follow fractures of weakened QCD (or subquark) forces,
and continue until all objects were small enough
to screen their own internal expansion. This would give the
early Universe a fractal structure, as for the mass
distribution of the ``final'' quark objects.
This is quite a different scenario compared
to the normal quark-nugget idea, because
the quark objects are formed directly from
the global plasma, and not as ``survivors''
inside a high-pressure hadron gas.

It is tempting to guess that this fractal
process led to the creation of more or less
separated regions containing a total
mass of typically galactic proportions
(or maybe those of galaxy-clusters). That
would explain why proto-galaxies were
formed so quickly after the Big Bang, a fact
that is hard to reproduce in simulations of a
purely gravitational development inside a
homogenous Universe of normal atomic
matter.

The smallest quark objects in such a fractal
distribution are obviously free nucleons,
while the most massive ones are harder to
define. One can think of a limitation given
either by stability criteria against a
gravitational collapse, or by relativistic
horizon arguments. Horizon arguments cannot
be strict since we do not know when the first
global split-up occurred. This problem was
discussed recently by Cottingham {\it et al.}
\cite{Cottingham94} within the context
of a field-theoretic model by Lee and Wick
\cite{Lee74}. The conclusion is that
supercooling of the global plasma promoted
a wider horizon than commonly believed,
which, in turn, would allow for quark
objects within a very wide mass range.
The maximal mass depends strongly
on the chosen values of model parameters,
and the authors give an example where
half the plasma hadronised, while the other
half remained as quark objects of up to
around one solar mass. This application of
the Lee-Wick model has been criticised recently
\cite{Kubis96}, and the horizon argument
is far from clear. We will therefore leave this
question open and discuss three different
scenarios, based on different mass scales for the
heaviest quark objects.

The most ``conventional'', and
therefore also least interesting,
such scenario would be that the primordial
matter was split up, immediately or in
successive steps, into rather small objects,
where some fraction quickly hadronised, while the
rest remained as stable quark objects.
The crucial size would be the few
dozen quarks hinted at by the MIT
bag model, and the hadronic fraction would be
the $10$ {\it per cent} or so given by
indirect observations of the dark-matter
content of galaxies. However, this simple-minded
scenario cannot explain why the
dark-matter quark objects would have a more
peripheral, and maybe spherical, distribution
in galaxies, while the luminous matter is mostly in
a central disc. Neither would it be credible
that a high fraction would be in the form of
massive machos of up to a solar mass, while
a substantial minority would be smaller
than a few dozen quarks. This would hardly
be reconcilable with a fractal structure, where
one expects the number of objects heavier than
a mass $M$ to drop like $M^{-\gamma}$,
with some power $\gamma$ of the order unity.

We therefore suggest that {\it all}
of the pregalactic matter was in the form of
rather massive quark objects, {\it i.e.} that only
an insignificant fraction hadronised immediately.
The nuclear matter in the
present-day Universe must then be the result
of an {\it enforced} phase transition of the primordial
quark matter, taking place only after
the first structuring into proto-galaxies.
Still, this hadronisation
must have started early enough for more
well-known processes to occur in due time,
{\it e.g.}, the forming of atoms, and the
decoupling of the cosmic microwave radiation.
Hence, also the latter process took place inside
proto-galaxies, explaining why there is still
a certain lumpiness in this (mostly) thermalised
radiation.

An enforced hadronisation can be a consequence
of an external or internal disturbance, leading to
an expansion (or split-up) of the object to a density
(or mass) well below that for hadronisation.
A simple cooling of the object due to Planck radiation
will not cause a hadronisation, once we assume that
quark matter remains the ground state also at $T = 0$.

There are now two interesting cases as for the choice
of an upper quark-object mass. One is that the
maximal mass was all from the start near
the upper limit for stability against
gravitational contraction (or collapse). The other is
that the very early quark objects indeed reached
galactic masses during a very brief epoch.
We will discuss these options separately.

\subsection{Stability criteria}

The most obvious criterion for the stability
of a massive (non-rotating) object is that its
radius must not fall below the Schwarzschild
radius given by

\begin{equation}
r_s = \frac{2M_{qm}G}{c^2},
\end{equation}

{\flushleft where} $M_{qm}$ is the mass.
For a quark-matter object this leads to a
critical (upper) radius for stability against
a gravitational collapse, $R_{crit}$, given by

\begin{equation}
R_{crit} = \sqrt {\frac{3c^2}{8 \pi \rho_{qm} G}},
\end{equation}

{\flushleft where} $\rho_{qm}$ is the density.
Assuming that it is, on the average, equal
to the density of a proton of radius $0.8$ fm,
this gives $\rho_{qm} = 3m_{p}/(4 \pi r^{3}_p)
\approx 10^{15}$ g/cm$^3$,
$R_{crit} \approx 14$ km and
$M_{crit} \approx 5M_{\odot}$.

However, a more refined analysis of a quark
object under internal gravitation shows that also
smaller systems would be unstable, since they
would slowly contract beyond the Schwarzschild
radius, and ultimately collapse into black holes.
The formalism for analysing such situations comprises
the so-called Tolman-Oppenheimer-Volkoff (TOV)
equations \cite{TOV39}, which are built on
general relativity, as well as on the equation of state of
a static, spherical, ideal-liquid system,
and on the assumption that stability is equivalent to
a zero-pressure at the surface of the object.
They are derived in many textbooks in general
relativity, {\it e.g.}, in \cite{Schutz90}, and
contain the following set of equations:

\begin{equation}
\frac{dp}{dr}=-\frac{[\epsilon (r)+p(r)][m(r)+4\pi r^3p(r)]}{r[r-2m(r)]},
\end{equation}

\begin{equation}
\frac{dm(r)}{dr}=4\pi r^2\epsilon (r),\hspace{1cm}m(r)=4\pi
\int_0^r\epsilon (r^{\prime })r^{\prime }{}^2dr^{\prime },
\end{equation}

\begin{equation}
p(r=0)=p_c,
\end{equation}

\begin{equation}
p(r=R)=0.
\end{equation}

{\flushleft Here} the TOV equations are expressed in
gravitational and natural units, $c=\hbar=G=1$.
$p$ is the total pressure, supposed to be built
up by a normal kinetic pressure, the external
vacuum pressure, the gravitational pressure and
an internal ``degeneration pressure'', which is a
phenomenological way of preventing the quarks
from violating the Pauli principle. $\epsilon$ is
the energy density and $m(r)$ is the mass inside
the radial coordinate $r$.

\begin{figure}
\centering
\epsfysize=6.5 cm
\leavevmode
\includegraphics{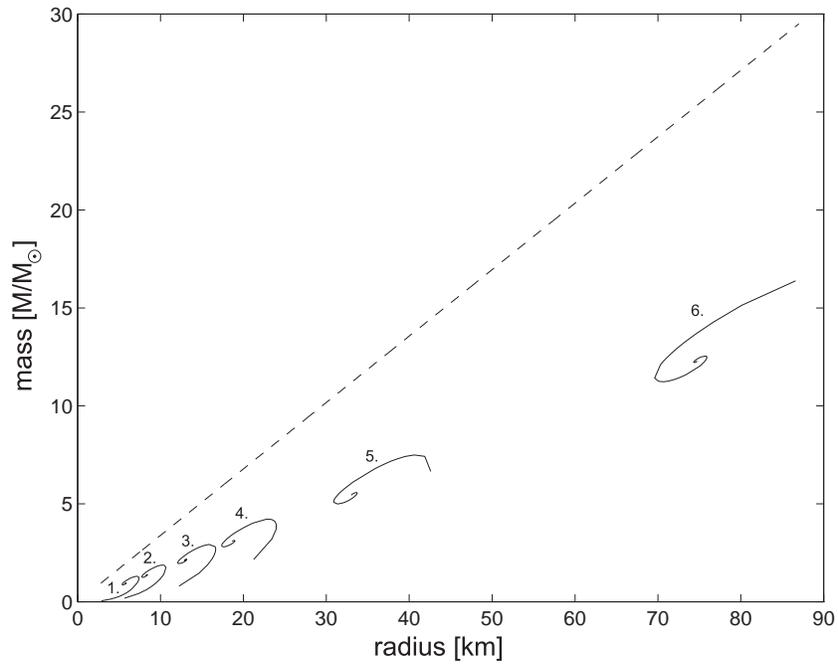}
%\epsffile{Darkmatterfig1.eps}
\caption{The stability relations (full lines) between
the mass and the radius of an MIT bag for different
values of the bag parameter $B$, namely
$B^{1/4}=180$ MeV (curve 1.), $150$ MeV (2.),
$120$ MeV (3.), $100$ MeV (4.), $75$ MeV
(5.) and $50$ MeV (6.). For clarity, only
the most relevant segments of the full
lines are shown. The hatched line shows
the criterion for collapse into a black hole,
as given by the Schwarzschild radius. Other parameter
values can be found in \cite{Enstrom97}.}
\label{massradius}
\end{figure}

Many such analyses of various stellar systems
can be found in the literature. They are based
on different equations of state, and other detailed
assumptions. Some recent computations by
one of us \cite{Enstrom97} give results that are
typical for analyses in the spirit of the MIT
bag model. A QCD-based equation of state for
a system of (equally many) $u$, $d$ and $s$ quarks
was used. Some of the results are illustrated
in Fig. \ref{massradius}.

For $B^{1/4} = 150$ MeV the
stable quark objects have masses below around
$\sim 1.8M_{\odot}$ and radii below $\sim 11$ km.
Similar results can be found in many analyses of
the stability of neutron stars, in particular those
with a sizeable quark-matter core \cite{Glendenning96}.
This applies also to rather different assumptions
about the internal structure and forces of such
quark or hybrid stars.

One example, which is of some
interest for our subsequent discussion, is
a quark object built up by {\it diquarks},
{\it i.e.}, tight pairs of two quarks,
generally prescribed to have total spin $0$.
Early suggestions of such systems were presented
by two of us in the late 1980s, including
discussions of their astrophysical relevance
\cite{Ekelin85,Ekelin86}.
A general review of the diquark
concept can be found in \cite{Anselmino93}, and
a topical review of their astrophysical relevance
is given in \cite{Fredriksson96}.
The interesting property of a
``diquark star'' is that the diquarks absorb
almost all attractive two-quark (QCD) forces
in the system, so that the net forces inbetween
diquarks are believed to be
either negligible, leading to a possible
Bose condensation \cite{Ekelin86},
or {\it repulsive}.

The latter situation was first analysed by
Donoghue and Sateesh \cite{Donoghue88},
and later applied to neutron-star cores
by Kastor and Traschen \cite{Kastor91}.
An interesting result is that a diquark star
is expected to be less dense than a
``conventional'' quark object. And with a
choice of somewhat extreme
(but not excluded) parameter values
it might even less dense than normal atomic nuclei.
Horvath {\it et al.} \cite{Horvath92}
refined this model, and concluded that diquark
stars might exist even without a
gravitational pressure. They also stressed
that the normal way of treating quark objects
within the MIT bag model must be revised if
there are diquarks in the plasma. All parameters
must, {\it e.g.}, be re-fitted to baryon
properties with diquark effects included,
in particular the bag parameter $B$.
As can be seen in Fig. \ref{massradius}
the mass-radius relation, and hence the
density, is quite sensitive to the $B$ value.

This particular example therefore
shows that although the maximal quark-object
mass is in the range of $1 - 5$ solar masses in
almost all detailed analyses, there is no general
agreement about such important details as the
average density of the objects. It should also be
stressed that the TOV equations with the external
pressure of the MIT bag model, and the equation
of state from QCD, does {\it not} exactly
reproduce the MIT bag model results for small
systems mentioned earlier. Hence, one cannot
trust the trend found in the TOV mechanism
that objects smaller than around one solar mass
would have a too low density for being quark
objects. This trend simply does not match the
MIT bag model result that objects with more
than a few dozen quarks are stable against
hadronisation.

In most quark-matter analyses of neutron
stars it is nevertheless assumed that
a density below $2 - 3$ times the normal
nuclear density is too low to allow for a
pure quark phase \cite{Glendenning96}.
Since the density in the surface layers
falls below these values even for the
most massive objects, it is often assumed
that a neutron star at most has a quark core,
surrounded by a crust of nuclear matter,
while smaller objects has no quark matter
at all. Even if this interpretation is correct,
the primordial quark objects suggested by
us need not turn into systems identical
to neutron stars. The reason is that these two
classes of dense objects were
created in completely different ways. Neutron
stars are the result of a compression of normal
atomic nuclei inside stars, while our quark
objects were created from the original quark-gluon
plasma soon after the Big Bang.

Neither is there any basic theoretical evidence
that quark matter must hadronise once its
density falls below some $2 - 3$ times that of
nuclear matter. These densities are the ones
believed to be needed for a phase transition
when nuclear matter is compressed at high
temperatures in heavy-ion collisions, and need
not have much relation to the conditions
needed for the reversed hadronisation of a
quark-gluon plasma with $T = 0$.
It seems more realistic to assume that a
quark object remains in a pure quark phase
until a nuclear density is reached as a consequence
of an enforced expansion.

We therefore conclude that there are no basic
theoretical arguments in contradiction to our main
assumption that the bulk of the dark matter
is identical to stable quark-matter
objects of a few solar masses and down.
One reservation must be made though,
namely for the fate of quark objects that
exceeded the mass limit of around $5M_{\odot}$
given by the Schwarzschild relation, or of the
$(1 - 5)M_{\odot}$ given by the TOV equations,
as discussed above. Such objects must have
turned into black holes, which could hence make
up a considerable fraction of the dark matter.
The literature on black holes is very
rich \cite{Luminet98},
and we will not go into details on the issue
of black holes as dark-matter candidates.

\subsection{Highly unstable quark-matter proto-galaxies}

It is tempting to speculate what would happen if
quark objects of up to galactic masses were
formed as an intermediate step in the very
early (fractal) split-up of the global quark-gluon plasma.
This would certainly avoid all problems with
the standard assumption of a gravitational accretion
into galaxies, and imply that galaxies are instead
shaped by {\it explosions}. The latter idea was
presented by Ambartsumian \cite{Ambartsumian58}
already in 1958 and cannot easily be
dismissed \cite{Harwit88}.
It turns out that galaxies are ``too often'' in
binary systems, which do not seem to be gravitationally
bound, but rather expanding. Also, there are
other more detailed paradoxes that might
be resolved within an explosive scenario.

We cannot prove that a fractal split-up after the Big Bang
ever gave rise to typical galactic masses, but we can
discuss some possible consequences.
Taking the Milky Way as an example, a typical
mass at this first stage was $M_{qm}
\approx 10^{12} M_{\odot} = 2 \times 10^{45}$ g,
assuming that the luminous matter now is around
$100$ billion solar masses, and that the total mass,
including dark matter, is an order of magnitude higher.

We also assume that this embryonic galaxy had a
density given by the MIT bag model, {\it i.e.},
in which it is almost independent of the mass,
and hence equal to that of the proton.
Therefore any quark-matter object is assumed
to have a density 
$\rho_{qm} \approx 10^{15}$ g/cm$^3$.
{\it If} the early quark-dominated Milky Way
was spherical, its radius
was hence given by $R_{qm} \approx 78,000$ km,
{\it i.e.}, around a quarter of a light-second.

Such an object would, naturally, be utterly
unstable, and it seems as if a quark-dominated
proto-galaxy would immediately fall into a
black hole. However, the situation is obviously
not that simple, because then the whole the
Universe would have done so at an earlier stage.
The situation is complicated by the fact that the
first formation of proto-galaxies took place in a
highly unstable and dynamically rapid sequence of events.

Two circumstances could have prevented the galaxy from
``disappearing'' instantly, namely its internal expansion
and its rotation, both presumably being extremely rapid
({\it i.e.}, relativistic). Such phenomena are
difficult to analyse within general relativity,
and we therefore limit ourselves, in this first
qualitative analysis to present some qualitatively
motivated conjectures about the very early phase of
the galaxy, including the creation of the first normal
hadronic matter.

It is likely that the expansion of the galaxy
was slowed down first of all in its central parts,
when the overall expansion of the
Universe, {\it i.e.}, of the ``coordinates'', was screened,
or balanced, by the internal gravitational and quantum
chromodynamic forces. Hence, the matter in the
central part would be the first to ``discover'' that the
Schwarzschild condition for collapse was fulfilled.
This information travelled at the speed of light, once the
proto-galaxy was formed, so that some fraction of the central
quark object must have collapsed in a fraction of a second,
and ``long'' before the whole galaxy had a chance to follow.
It is well known that many galaxies have central black holes,
with masses of $(10^{6}-10^{10})M_{\odot}$
\cite{Franceschini98}, which, in our model,
could be from this very early phase.

The central matter that collapsed into such a
black hole, ``emptied'' a sphere in the galaxy,
with a radius of around $4,000$ km, assuming a typical
mass of $10^{8}M_{\odot}$ (note that
the central black hole in the Milky Way is probably
much smaller; of the order of $10^{6}M_{\odot}$).
The original ``MIT bag'' therefore
turned into a hollow sphere, and this disruption
might well have caused an internal explosion
that ripped the proto-galaxy apart into the
smaller quark objects discussed in the previous
section. Some details of this will be discussed
in the next chapter, since it might be connected to
a central phase transition into normal matter.

The idea that a central explosion broke up the
whole proto-galaxy before it had the chance to
know that it was smaller than its own Schwarzschild
radius does not need to be a paradox. The expansion
and rotation would possibly have given the same
results by themselves, and galaxies could therefore
lack a central black hole.

Assuming that the rotation of a typical spiral galaxy
is primordial, one can estimate the original angular
momentum inside the quark object. Strangely enough,
it seems as if the galactic halo has a negligible
net angular momentum, since there is no correlated
motion among halo stars. The disc, however, has an
angular momentum of the order of $10^{72}$ erg s,
which hints at a relativistic rotation in the inner part
of the quark-dominated proto-galaxy. This perhaps
indicates that the disc-form is closely connected to
the Kerr metric of a rotating central black hole, while
the more spherical halo is related to the more
symmetric primordial expansion and early explosion.

Apart from the central black hole and possible
phase transition caused by it, the explosion broke
up the outer part of the quark-gluon plasma
into smaller objects, which either continued to
break up, fell into small black holes, or
stabilised in the sense discussed in the
previous section. This means that the
subsequent history of the early galaxy would be
much the same in the two scenarios, differing
only as for the structure of the galactic centre.

\section{Hadronisation and gamma-ray bursts}

Here we suggest that the normal nuclear matter
was created ``violently'' in the proto-galaxy, and
mostly in its centre. Such events gave rise also
to bursts of gamma radiation, some of which
can still be observed from far-away galaxies
in the form of the much-discussed gamma-ray bursts.

We start by discussing some general aspects of such
bursts, and then turn to the two different sources
of bursts that are connected to the two scenarios
in the previous chapter. A preliminary discussion
was presented in \cite{Anoushirvani97}.

\subsection{Gamma-ray bursts and distance scales}

The bursts of intense gamma rays (GRB), first
observed in the 1960s by the Vela military satellites
and disclosed to the civilian research
community in 1973 \cite{Klebesadel73},
have confounded physicists and astronomers ever
since. Although the outbursts must be very energetic,
the actual value of the total energy depends on
their distance from the earth. The time-span of the
bursts is typically $0.01-1000$ s, and no
characteristic features, such as
spectral lines have been detected, with two
interesting exceptions.
The burst named GRB970508 has been related to
an object that appeared as an optical transient
shortly after the burst, revealing clear spectral
absorption lines. The absorbing body, which can be
either a host-galaxy of the GRB,
or an intervening foreground body, has been shown
to have a redshift parameter $z \approx 0.835$.
Hence the GRB source itself has $z \geq 0.835$.
There is also some indirect evidence for an
upper limit, $z \leq 2.3$ \cite{Metzger97}.
Similar values have been suggested for the
more recent GRB971214 \cite{Reichart98},
namely $1.89 \leq z \leq 2.5$.

There are still no clues to whether these GRBs are
``average'' in any sense. Future optical GRB
transients with spectral lines is certainly needed
before a distance-scale can be confirmed. Only
after such a scale has been determined, will
it be possible to discriminate between the several
dozen published theoretical models of the origin
of bursts. It should be noted that all other efforts
by authors of GRB publications to pinpoint an
absolute distance to a particular GRB, or a well-chosen
class of GRBs, are model-dependent, and therefore
less reliable. Such estimates seem to
cluster around $z$ values of $1 - 2$.

In the 1980s, the consensus among researchers was
that the bursts originate within our own galaxy
\cite{Harding91}. When the Burst and Transient Source
Experiment (BATSE) \cite{Meegan96}, aboard the
Compton Gamma Ray Observatory (CGRO), began
to produce much more data it became evident
that the gamma-ray bursts are distributed
isotropically in the sky, not following the visible
outlines of the Milky Way (nor of the Andromeda).
The opinion among astrophysicists then swayed
to models assuming a cosmological origin. A few
thousand gamma-ray bursts have been detected
to date (and there have been roughly as many
different publications on the subject).

The most popular GRB model seems to be that
they originate from the binary collapse of two very
compact star remnants; neutron stars, black holes,
or a combination thereof. Such models take it for
granted that these events occur at random in all
normal galaxies, typically $1 - 100$
per a million years
per galaxy. The rarity of such mergers would explain
why none of the detected bursts has yet occurred close to
a visible galaxy, and why there has been no repetition
of events from the same locations. The low frequency
is also in line with estimates of the number of
neutron stars in galaxies, and even the energy release
seems to fit what would be expected if two neutron
stars merge. If the GRBs are evenly distributed
among galaxies, the bursts seem to release
$10^{51} - 10^{53}$ erg of gamma rays,
which is one or two orders of magnitude less
than the expected gain in gravitational
energy due to a merger.
A weak point of all such models is, however,
that it is hard to understand how this gravitational
energy converts into gamma rays. A chain of
processes has been suggested, where the
primary energy turns into a shock-wave of neutrinos,
which annihilate into $e^+e^-$ pairs, and ultimately
into gammas when the charged leptons hit
the thin interstellar medium.

In our model GRBs are a consequence
of the phase transition of quark matter into
hadrons in the early proto-galaxies,
and hence we assume that they are not at
all evenly distributed in
space, with a universal frequency per galaxy, but
instead strongly biased toward large distances,
{\it i.e.}, the early Universe and high redshifts.
Since such a phase transition would give
a wealth of almost directly produced gammas,
there will at least not be any problems with
understanding the very high gamma intensities.

Lacking an absolute distance-scale, it is, in fact,
almost trivial to fit a distribution of GRB distances,
with any chosen average distance, to the observations
of gamma-ray fluxes. We will demonstrate how this
works, with a simple choice of such distributions.

As is customary, we restrict ourselves to an
Einstein-de Sitter universe with vanishing
cosmological constant and global curvature.
This choice seems, by comparison to observational
data, to be a reasonably good approximation of the
Universe. We also assume that the individual bursts
can be treated as ``standard candles'', {\it i.e.},
that the characteristics of a typical (``average'')
burst stays the
same during the full burst epoch
and in all galaxies. This condition need not
be true and can easily be relaxed, but that
would not give any deeper insight into
the relevant processes.

Each burst is assumed to emit the radiation
uniformly in all directions ({\it i.e.}, not in
beams). Relaxing this condition would, of course,
require more bursts, and a lower energy
release per burst. As will be argued later
there are indeed reasons to believe that
there is some beaming in phase transitions
caused by the merging of two quark objects.

There seems to be no consensus regarding
possible time-dilation effects in GRB spectra, nor
regarding an intrinsic duration-luminosity correlation
(incompatible with the standard-candle assumption),
with strong bursts having shorter duration and
{\it vice versa}. We simply ignore such
(presumably weak) effects in the following analysis,
and concentrate on the number/peak-flux relation.

Taking one or more of these complications into account
would not change our general observation that a wide
range of GRB space distributions can be fitted
to the flux data.

The flux of a particular gamma-ray burst can, if the
conditions mentioned above are satisfied, be given as
a function of its redshift parameter, $z$, \cite{Misner73}

\begin{equation}
P(z) = \frac{L(z)}{4 \pi r(z)^2 (1 + z)^2},
\end{equation}

{\flushleft where} $L(z)$ is the luminosity of the
burst. Note that the ``redshift'' is normally
defined as ``$1+z$''. The present distance to the source, $r(z)$,
depends on the cosmological model. In our case
(flat Einstein- de Sitter space), this relation reads

\begin{equation}
r(z) = \frac{2c}{H_0} (1 - \frac{1}{\sqrt{1 + z}}),
\end{equation}

{\flushleft where} $H_0$ is the Hubble constant
(taken as 75 km/s/Mpc).

The source luminosity detectable by an instrument
near the earth, with an effective energy detection
window between $E_{min}$ and $E_{max}$,
is given by

\begin{equation}
L(z) = \int_{E_{min} (1 + z)}^{E_{max}
(1 + z)} \phi (E) dE,
\end{equation}

{\flushleft where}

\begin{equation}
\phi (E) = A_0 \frac{e^{-E/kT}}{E}
\end{equation}

{\flushleft is} the spectral form (thermal
bremsstrahlung) conventionally chosen
for modelling the burst \cite{Horack96}.
$kT$ is a characteristic energy for a typical
burst, chosen to be $350$ keV.
For BATSE, $E_{min} = 50$ keV and
$E_{max} = 300$ keV.

For simplicity, we assume that the number
density, $\rho (r)$, of the bursts is a gaussian,

\begin{equation}
\rho = C \, \frac{1}{\sqrt{2 \pi} \sigma}
e^{-(r - r_0)^2 / 2 \sigma^2},
\end{equation}

{\flushleft centred} around $r_0 = r(z_0)$,
and with variance $\sigma$. The normalising
constant $C$ is fitted to the data.
A homogeneous distribution in Euclidean
(fairly nearby) space, within a spherical shell
with nothing outside, is also compatible with 
the BATSE data within observational errors,
although such an abrupt cut-off seems unphysical.
A smoothed-out version of such a distribution,
or some completely different distribution altogether,
could equally well be fitted to the data. For brevity,
we only consider a gaussian distribution here.

Finally, in a given concentric spherical shell, there will
be a differential number of bursts given by

\begin{equation}
n = \frac{dN}{dr} = 4 \pi \, \rho \, r^2,
\end{equation}

{\flushleft where} $N$ is the total number of GRBs
within a sphere of radius $r$, {\it i.e.}, the total number
with observed fluxes below some value (because of our
standard-candle assumption).

\begin{figure}
\centering
\epsfysize=6.5 cm
\leavevmode
\epsffile{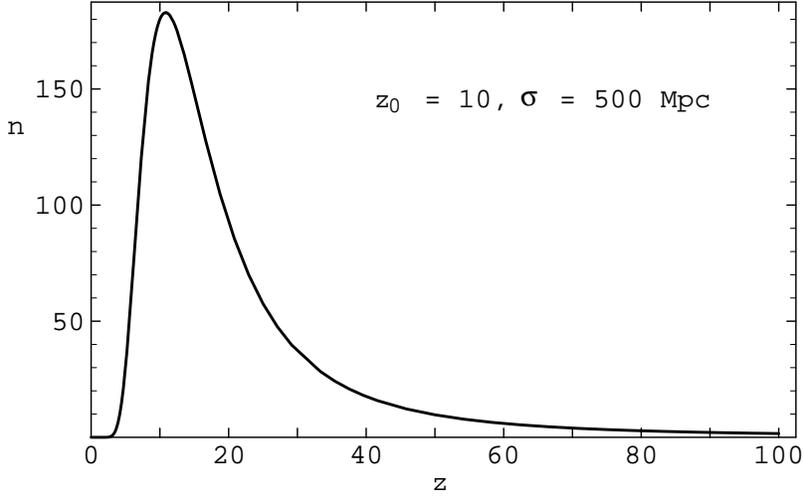}
\caption{An example of the differential number
($n$) of gamma-ray bursts as a function of redshift
parameter ($z$) in flat Einstein-de Sitter space
for a gaussian
distribution with the quoted values of
$z_0$ and $\sigma$. The form is chosen as to
fit the BATSE data on the differential
number distribution as a function of apparent
intensity.}
\label{nz}
\end{figure}

\begin{figure}
\centering
\epsfysize=6.5 cm
\leavevmode
\epsffile{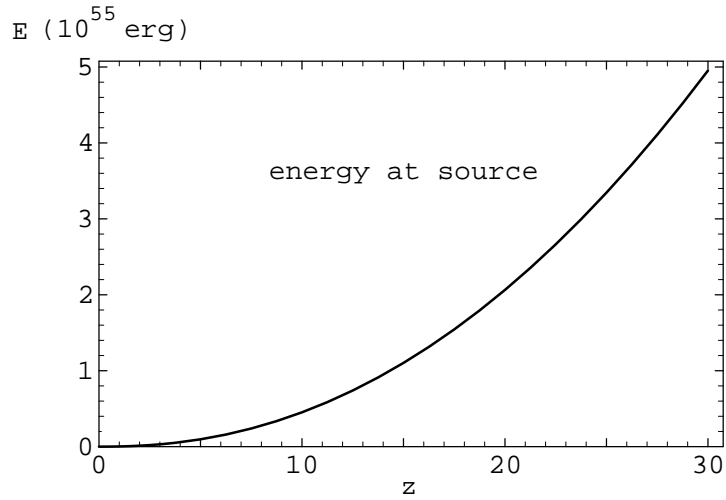}
\caption{The total gamma-ray energy emitted
by the source, given a flat Einstein-de Sitter space,
and a typical detected energy flow of
$10^{-5}$ erg/cm$^2$. The Hubble constant,
$H_0$, is taken as $75$ km/s/Mpc.}
\label{rz}
\end{figure}

It is, in principle, possible to fit the BATSE data on
the $n$ distribution reasonably
well with any choice of $z_0$, while $\sigma$
would be derived by the fit
(being smaller for higher $z_0$ value).
A typical form of $n$ as a function of $z$
is shown in Fig. \ref{nz}. The gamma-ray
energy emitted by a ``standard-candle'' source,
which is needed to fit the typical energy flow
into the detector ({\it i.e.} around $10^{-5}$
erg/cm$^2$) is shown as a function of $z$ in
Fig. \ref{rz}.

We conclude this section with the observation that
there is no real limitation on the average distance
to the bulk of gamma-ray sources, except for hints
from two events (GRB970228 and
GRB971214) with redshift $z$ values estimated
to be $0.835 - 2.3$ and $1.89 - 2.5$, where
the upper bounds are motivated mainly by
``non-observations'' (of redshifted Lyman
absorption lines in hydrogen).

\subsection{A central galactic explosion}

If indeed a supermassive black hole was created
in the centre of the very early and unstable
quark-matter proto-galaxy, there should have
been an immediate implosion of the plasma
into the ``vacuum'' dug out at the centre.

This is a type of enforced expansion
of the quark matter that could have catalysed
a phase transition into normal hadronic matter.
It would {\it not} be equivalent to a condensation
outwards on the periphery of the bag, because the
physical situation on the inner and outer surfaces
are entirely different. On the outer surface there
is still gravity and confinement
(vacuum pressure), while on the inner surface neither of
these would prevent the plasma from an implosion.
Hence the galaxy would undergo a phase transition from
inside out, which explains why normal matter,
unlike dark matter, is in the central parts of a galaxy
(and, of course, also why there are black holes in
so many galactic centres).

A black hole of mass $10^{8}M_{\odot}$ would
need to swallow quark matter within an original
radius of roughly $3,000$ km.
If this was subsequently filled with hadronising matter, it
could sustain a mass of around $3 \times 10^{7}M_{\odot}$
without exceeding normal nuclear densities.
For simplicity, we ignore that some of that matter would
also fall into the black hole, and so on. If we now
assume that this enforced
phase transition released some $10 - 100$ MeV of energy
for each produced nucleon, the total energy radiating
from the centre of the proto-galaxy would be of the
order of $10^{60} - 10^{61}$ erg. If, for some
reason, all the current luminous matter was
created in this implosion, the corresponding
energy (for the Milky Way) would be roughly
$10^{64}$ erg.

Such energies are
far in excess of the gravitational binding energy
of a quark object of $10^{12}$ solar masses, and would
hence suffice to tear the galaxy apart and provide a
sizeable outflow of energetic hadrons and gammas.
The effect would naturally be even stronger if
the proto-galaxy was already highly unstable,
and even diluted (supercooled) at the centre, due
to the overall expansion of the Universe and an internal
rotation.

It is interesting to note that the emitted gamma
energy of a typical gamma-ray burst would
correspond to around $10^{61}$ erg,
would the source be at a
redshift with $z_0 = 10000$.
Such $z$ values are typical for the epoch of
the global quark-hadron phase transition in the
conventional Big Bang scenario.

Hence we have sketched a mechanism for creating,
at a very early stage, a massive black hole and a
sizeable amount of normal matter in the centre
of the galaxy, while pushing the dark matter to
the periphery in an explosive event.
All this would be marked with
an enormously energetic gamma-ray burst. The
fact that normal matter is in disc-form might
be a consequence of an original rotation of the
galactic centre, while the sphericity of the dark-matter
halo might mirror the large-scale isotropy of the
burst (the gammas would radiate at random,
in spite of the rotation).

It is tempting to speculate that these gamma-ray
bursts are still visible, since it would certainly
be thrilling if we can still see the
creation of whole galaxies. This idea would probably
not be in contradiction with such well-known
GRB features as their overall time-scale and
``spiky'' time development. It would, however,
be hard to understand how they can still appear
as pointlike events, regarding the fact that the
proto-galaxies must have been formed during
an epoch when they were closely packed
in space. The gammas should then have been 
efficiently scattered and absorbed by nearby
matter, and hence thermalised quickly.

Also, it is unlikely that {\it all} atomic matter
in a galaxy was created through an inflow
of quarks in the limited space created by the
black hole. We therefore assume that the bulk of
normal matter was created somewhat later, from
the remaining, smaller quark objects in the galaxy.

\subsection{Binary quark-object mergers}

Here we assume that the very early proto-galaxy
was a collection of quark objects of all
sizes allowed by the stability criteria discussed above.
We do not discriminate between the two
possible prehistories, {\it i.e.}, whether this
epoch was preceded by a central galactic
explosion as described above, or a more peaceful
gathering of these objects into galactic regions.

The most credible way of enforcing a
phase transition into normal matter in
the galactic centre would then be through
{\it mergers in quark-object binaries}.
Again, we suggest that such transitions
would also produce gamma-ray bursts
and that these constitute the bulk of the
observed GRBs.

It is worth noting that this idea is similar,
but still orthogonal, to the popular conjecture
that gamma-ray bursts originate from
neutron-star mergers \cite{Blinnikov84}.
A majority of such models rely on some
either unspecified, or very complicated,
mechanism for converting
the energy-gain into gammas. However,
there are also models were the gamma-ray
bursts originate from a hadron-quark phase
transition inside a neutron star, or in
connection to a merger \cite{Cheng96}.
These transitions are {\it from} a hadron
phase {\it to} a quark phase, and hence
opposite to our scenario.

A typical sequence of events inside a single
neutron star (pulsar) is that its rapid rotation
is gradually slowed down by gravitational
radiation (and maybe also ``frame dragging'').
This leads to a weakening of the internal
centrifugal forces, and hence an increase
of the central density, until (possibly) a phase
transition occurs, creating a quark-matter core.

In our model, a quark-object binary would,
on the contrary, spiral into a closer and closer
orbit, with an increasing rotational frequency,
until the two objects would trigger a mutual
phase transition into hadronic matter. This could
take place as a consequence of two different
physical phenomena.
One would be that there would form a bridge
of matter between the two objects once they
are close enough, {\it i.e.}, when the confinement
in one object would be counterbalanced by the
gravitational pull from the other. Such matter
flows occur in many cosmic
situations, {\it e.g.}, between galaxies, 
between normal stars in a binary system, and
from an accretion disc into a black hole.
There are rather well-established formalisms
for analysing such situations \cite{Carroll96}.
Another catalyser
could be that the tidal effects, and the centrifugal
forces, in the last stage of the merging, would
pull one or both objects into such a low density
that they would hadronise separately.

Suppose that such mergers would
lead to the hadronisation of half the
matter in the two quark objects.
Then the energy release from them would
be maximally $10^{53} - 10^{54}$ erg, assuming that
a maximal mass of $5M_{\odot}$ would be
involved, and that the
microscopic energy release is $10 - 100$ MeV per
produced nucleon. If a sizeable fraction of this energy
goes directly into gammas, we would get
a gamma-ray burst typical for source redshifts with
$z = 5 - 10$, {\it i.e.}, before the epoch of star
formation in galaxies.

One might ask whether such a scenario would
be in jeopardy with some obvious observational
restrictions. Although this new model for gamma-ray
bursts has many macroscopic and microscopic
features in common with GRB models with neutron-star
mergers, there are two important differences. Firstly,
we do have any problems to understand why so many
gammas can result from a binary merger, since they
come directly from the source, and not from the
interstellar medium. Secondly, our mergers must have
been much more frequent in the past than those of
neutron stars, since they involve the bulk of galactic
(dark) matter, and are the source of the
normal (luminous) matter.

Neutron-star mergers are estimated to happen at a
frequency of roughly $10^{-4} - 10^{-6}$ y$^{-1}$
during the life of a galaxy. Quark-object mergers
must, however, have occurred {\it at least} around
$2 \times 10^{10}$ times in the Milky Way,
assuming that all the luminous matter of
mass $10^{11}M_{\odot}$ was
created in mergers where up to $5M_{\odot}$
hadronised at a time. This gives an average
frequency of more than one GRB per year per
typical galaxy. An overwhelming majority of
these must originate in the galactic centre,
where the fraction of binaries must have
been much higher than in the halo, since the
tendency to form binaries depends strongly
on the overall density. This explains why
the normal matter is concentrated
in the galactic centre.

Such a high frequency is naturally excluded
by the clear lack of GRB repeaters
from correlated regions in space. In our model
this ``paradox'' can be explained by several
simple facts, of which three will be discussed below.

{\it Firstly}, a majority of the mergers must have happened
in the very early, quark-matter dominated galaxy.
We know that the bulk of atomic matter was created
shortly (maybe only a hundred thousand years) after
the Big Bang. The galaxy was then very compact
(according to our model), and the radiation pressure
from all the early mergers can have helped the galactic
halo to an early expansion. In this dense environment, the
bursts could probably not be seen from outside the
galaxy. Rather, the bursts contributed to the general,
thermalised radiation background, which was later
decoupled from the atoms and still exists as the
microwave background.
Such an early self-destruction of quark objects
led to a lack of binaries in the
current dark matter in low-redshift
galaxies, including our own.
There is, for instance, a ``disturbing''
deficit of binary events among the
observed machos, since only one macho has been
observed to have a double-spiked structure.
Contrary to the situation for neutron
stars there is no production of new
quark bodies, and new binaries can
come about only through random gravitational
captures.

{\it Secondly}, only the most violent mergers of
quark objects would give gamma
energies in the range of $10^{53}$ erg. The
bulk of the events might be orders of magnitude
less energetic, and maybe even unobservable if they
occur mainly in galactic centres. The ultimate
test of this idea would of course be to observe
one such event in our own galaxy, with clear
indications of a quark-hadron phase transition.

{\it Thirdly}, a gamma-ray burst from a hadronising
bridge of matter between two nearby objects can be
strongly ``beamed'', and hence less energetic
than what is expected for an isotropic
emitter. In fact, the radiation should
be more or less compressed into a disc-like
region perpendicular to the symmetry line
between the centres of the two objects. This is so,
because the phase transition should start only when
the two quark objects come very close to each
other (say, within a few km).
Then they would simply shadow
the gammas in directions away from such
a central disc.
The whole system is also rotating very rapidly
at this stage, maybe with a period of a fraction
of a second, so that the disc-like radiation zone
would sweep the sky rapidly.

The latter mechanism, {\it i.e.} a disc-like
radiation zone, rotating rapidly around an
axis in its own plane (and maybe even wobbling),
could be the true cause
of the very spiky nature of many gamma-ray
bursts. Another explanation of these spikes
is that the matter does not flow smoothly
into the hadronising region, but as several
jets, or in the form of chunks of quark-gluon
plasma.

The very irregular time structure of a GRB
can also be caused by a range of other complications;
one being the reheating of the original quark
matter by the radiation from the first stage of
the phase transition, and perhaps also an instant
radiation pressure on the objects, which pushes
them into a wider orbit for a short time, until
they come closer again, and the transition
continues.

As for the total time-scale of $0.1 - 1000$ s
for a GRB, it does not seem unlikely
that this would be typical for a partial or
complete phase transition of quark objects of
up to a few solar masses. The same conclusion
can be drawn for neutron-star mergers, and
the time-scale is probably set by the size of
the objects and the time it takes for the excess
energy to leave the region.

The energy distribution of the gammas is
known to be non-thermal, and with a broad
and smooth distribution around a few hundred
keV. This is expected
for gammas created inside the source, and
with a partial thermalisation in a dense ``nuclear''
environment. The thermalisation cannot possibly
be complete, because the region is only a few km
thick, and it should also blow apart due
to the internal radiation.

Much of the current work on GRBs focuses
on the observed X-ray afterglows from six
of them, three of which have been correlated
also to optical afterglows, the most recent one
being the GRB971214 \cite{Heise97,Antonelli97}.
In our model, the hadronised matter
most certainly emits radiation of gradually
decreasing frequencies, as a consequence of
the further development of the created
nuclear and atomic matter. The first
phenomenon would be the weak decays of the
$s$ quarks into $u$ and $d$ quarks (in roughly
$10^{-10}$ s), and the following beta decays
of the neutrons in a few thousand seconds,
leading to a shower of neutrinos and electrons
(roughly $10^{57}$ from a solar-mass object).

The electrons would then slow down, either
inside the cloud or when hitting the interstellar
medium, and give off X rays. An optical
afterglow can be a consequence of electrons
being captured by the protons from neutron
decay, and hence mark the creation of hydrogen
atoms.

All these processes would be a rapid
mini-version of the standard Big Bang scenario,
where the much shorter time-scale is
determined by the size of the object. The
kilometre-sized region of hadronised matter
simply cools off very rapidly after
the re-heating caused by the phase transition.

We therefore conclude this discussion of gamma-ray
bursts by pointing out that there are no observational
data that contradict the idea that GRBs come from
phase transitions of massive quark objects. On the
contrary, this model avoids some of the problems
of models built on neutron-star mergers.

\section{Conclusions}

The model presented here is, to the best of our knowledge,
the first one that relates gamma-ray bursts to the
dark-matter problem, and to the creation of normal
atomic matter in the galactic centres.

These ideas limit the model in the sense that they
require a completely different sequence of events
immediately after the Big Bang, as compared to the
standard scenario. The most original ``details'' here are
that the global quark-gluon plasma must have gathered into
proto-galaxies {\it before} the quark-hadron phase transition,
and that the final structure of the galaxy can be a result
of an explosive development, following the creation
of a massive, central black hole, or a fireworks
of early mergers of quark-object binaries.

Such a fantastic scenario does not seem to
violate qualitative astrophysical facts, but
certainly most of the more conventional models,
including the later phases of the so-called
standard Big Bang scenario. It remains to analyse
if it is also in line with the wealth of more
detailed data from the cosmos. In the longer
perspective it will also be interesting
to wait for some clear-cut signatures of the
more than $1000$ billion massive quark
objects that we expect to orbit our own galaxy
{\it e.g.}, in the form of a much better statistics
in the studies of gravitational micro-lenses.

This project is supported by the
European Commission under contract
CHRX-CT94-0450, within the network
"The Fundamental Structure of Matter".

\end{document}